\documentstyle[aps,aipbook,floats]{revtex}

\newcommand{\bean}{\begin{eqnarray*}}
\newcommand{\eean}{\end{eqnarray*}}
\newcommand{\hPi}{\mbox{$\widehat{\Pi}$}}
\newcommand{\hGa}{\mbox{$\widehat{\Gamma}$}}

\newcommand{\eg}{\mbox{$\gamma$}}

\def\be{\begin{equation}}
\def\ee{\end{equation}}
\def\bea{\begin{eqnarray}}
\def\eea{\end{eqnarray}}
\def \dsl {\partial \kern-.55em{/}}
\def \Dsl {D \kern-.65em{/}}
\def \qsl {q \kern-.45em{/}}
\def \slp {p \kern-.45em{/}}
\def \ksl {k \kern-.45em{/}}

\def \g5 {\gamma _5}

\begin{document}
\title{Standard Model Higher Order \\ Corrections  to the $WW\gamma / WWZ$
 Vertex}

\author{Joannis Papavassiliou}
\address{New York University, New York 10003, USA}

\maketitle

\begin{abstract}
Using the S--matrix pinch technique we obtain to one loop order
gauge independent $\gamma W^-W^+$ and  $Z W^-W^+$ vertices in the context of
the standard model, with all incoming momenta off--shell. We show that the
vertices so constructed satisfy simple QED--like Ward identities.
These gauge invariant vertices give rise to
expressions for the
magnetic dipole and electric quadrupole
form factors of the $W$ gauge boson,
which, unlike previous treatments,
satisfy the crucial properties of infrared finiteness and
perturbative unitarity.
\footnote{To appear in the proceedings of the
International Symposium on Vector Boson Self Interactions, Feb. 1-3, 1995
UCLA.}
\end{abstract}

\section*{Introduction}

A new and largely unexplored frontier on which the ongoing search for new
physics will soon focus is the study of the structure of the three-boson
couplings \cite{Fawzi}.
A general parametrization of the trilinear gauge boson vertex for
two on--shell $W$s and one off--shell  $V=\gamma,~Z$ is \cite{Bardeen}
\bea
\Gamma_{\mu\alpha\beta}^{V}= &-ig_{V}~\Big[~
 ~f \left[~ 2g_{\alpha\beta}\Delta_{\mu}+ 4(g_{\alpha\mu}Q_{\beta}-
g_{\beta\mu}Q_{\alpha})~\right]\nonumber \\
& +~ 2\Delta\kappa_{V}~(g_{\alpha\mu}Q_{\beta}-g_{\beta\mu}Q_{\alpha})
\nonumber \\
& ~~~~~~~~~~~~~~+~  4\frac{
{\Delta Q_{V}}}{{M_{W}^{2}}}
(\Delta_{\mu}Q_{\alpha}Q_{\beta}-
\frac{1}{2}Q^{2}g_{\alpha\beta}\Delta_{\mu})
{}~~\Big]~,
\eea
with $g_{\eg}= gs$, $g_{Z}= gc$, where $g$ is the $SU(2)$
gauge coupling, $s\equiv sin\theta_{W}$ and $c\equiv cos\theta_{W}$.
In the above formula terms which are odd
under the individual discrete symmetries of C, P, or T
have been omitted.
 The four-momenta $Q$ and $\Delta$
 are related to the incoming momenta $q$, $p_{1}$ and $p_{2}$ of
the gauge bosons $V,~W^-$and $W^+$ respectively, by
$q=2Q$, $p_{1}=\Delta -Q$ and $p_{2}=-\Delta - Q$~.
The form factors $\Delta\kappa_{V}$ and $\Delta Q_{V}$,
also defined as
\be
\Delta\kappa_{V}= \kappa_{V} + \lambda_{V} - 1 ~,
\ee
and
\be
 \Delta Q_{V}= -2\lambda_{V} ~,
\ee
are compatible with C, P,
and T invariance, and
are  related to the magnetic dipole moment $\mu_{W}$ and the electric
quadrupole moment $Q_{W}$, by the following expressions
\cite{Goun1},\cite{Hagi},\cite{Z1},\cite{Z2}:
\be
\mu_{W} = \frac{e}{2M_{W}}(2+ \Delta\kappa_{\gamma})~,
\label{dipole}
\ee
and
\be
Q_{W}= -\frac{e}{M^{2}_{W}}(1+\Delta\kappa_{\gamma}+\Delta Q_{\gamma})~.
\label{quad}
\ee
In the context of the standard model,
their canonical, tree level values, are
$f=1$ and  $\Delta\kappa_{V}=\Delta Q_{V}=0$.
To determine the radiative corrections to these quantities
one must cast the resulting one--loop expressions in the following form:
\be
\Gamma_{\mu\alpha\beta}^{V}= -ig_{V}[
 a_{1}^{V}g_{\alpha\beta}\Delta_{\mu}+ a_{2}^{V}(g_{\alpha\mu}Q_{\beta}-
g_{\beta\mu}Q_{\alpha})
 + a_{3}^{V}\Delta_{\mu}Q_{\alpha}Q_{\beta}]~,
\label{1loopParametrization}
\ee
where $a_{1}^{V}$, $a_{2}^{V}$, and $a_{3}^{V}$ are
 complicated functions of the
momentum transfer $Q^2$, and the masses of the particles
appearing in the loops.
It then follows that  $\Delta\kappa_{V}$ and $\Delta Q_{V}$
are given by the
following expressions:
\be
\Delta\kappa_{V}=\frac{1}{2}(a_{2}^{V}-2a_{1}^{V}-Q^{2}a_{3}^{V})
\label{1loopdeltakappa}
\ee
and
\be
\Delta Q_{V}= \frac{M^{2}_{W}}{4}a_{3}^{V}~.
\label{1loopdeltaQ}
\ee
Calculating the one-loop expressions for $\Delta\kappa_{V}$ and
 $\Delta Q_{V}$ is a non-trivial task, both from
the technical and the conceptual point of view.
If one
calculates
just the Feynman diagrams contributing to the $\gamma W^{+}W^{-}$
vertex and then extracts from them the contributions to
$\Delta\kappa_{\gamma}$ and $\Delta Q_{\gamma}$,
one arrives at
expressions that are
plagued with several pathologies, gauge-dependence being one of them.
Indeed, even if the two $W$ are
considered to be on shell, since the incoming
photon is not, there is no {\sl a priori}
 reason why a gauge-independent answer
should emerge. In the context of the renormalizable $R_{\xi}$ gauges
the final answer depends on the
choice of the gauge fixing parameter $\xi$, which enters into the one-loop
calculations through the gauge-boson propagators
( $W$, $Z$, $\gamma$, and unphysical scalar particles).
In addition, as shown by an explicit calculation
performed
in the Feynman gauge ($\xi=1$), the answer for
$\Delta\kappa_{\gamma}$
is {\sl infrared divergent} and
violates perturbative unitarity,
e.g. it grows monotonically for $Q^2 \rightarrow \infty$
{}~\cite{Lahanas}.
 Clearly, regardless of the measurability of
quantities like $\Delta\kappa_{\gamma}$ and $\Delta Q_{\gamma}$, from
the theoretical point of view one should at least be able to satisfy
such crucial requirements as gauge-independence and infrared finiteness,
when calculating the model's prediction for them.
Indeed, all the above pathologies may be circumvented
if one adopts the pinch technique (PT), first invented by Cornwall
\cite{{CornMarseille}}.
The application of this method gives rise to
new expressions,
$\hat{\Delta}\kappa_{\gamma}$ and $\hat{\Delta} Q_{\gamma}$,
which are
gauge fixing parameter ($\xi$) independent,
ultraviolet {\sl and} infrared finite, and
well behaved for large momentum transfers $Q^{2}$~\cite{Kostas}.

\section{The pinch technique}
The simplest example that demonstrates how the PT works is the gluon
two point function \cite{Bib}.
Consider the $S$-matrix
element $T$ for the elastic scattering
such as $q_{1}{\bar{q}}_{2}\rightarrow q_{1}{\bar{q}}_{2}$,
where $q_{1}$,$q_{2}$ are two
on-shell test quarks with masses
$m_{1}$ and $m_{2}$.
To any order in perturbation
theory $T$ is independent
of the gauge fixing parameter $\xi$.
On the other hand, as an explicit calculation shows,
the conventionally defined proper self-energy
depends on $\xi$. At the one loop level this dependence is canceled by
contributions from other graphs, which,
at first glance, do not seem to be
propagator-like.
That this cancellation must occur and can be employed to define a
gauge-independent self-energy, is evident from the decomposition:
\begin{equation}
T(s,t,m_{1},m_{2})= T_{0}(t,\xi) +
\sum_{i=1}^{2}T_{i}(t,m_{i},\xi)
+T_{3}(s,t,m_{1},m_{2},\xi)~,
\label{S-matrix}
\end{equation}
where the function $T_{0}(t,\xi)$ depends
kinematically only on the Mandelstam variable
$t=-({\hat{p}}_{1}-p_{1})^{2}=-q^2$,
 and not on $s=(p_{1}+p_{2})^{2}$ or on the
external masses.
Typically, self-energy, vertex, and box diagrams
contribute to $T_{0}$, $T_{1}$, $T_{2}$, and $T_{3}$, respectively.
Such contributions are $\xi$ dependent, in general. However, as the sum
$T(s,t,m_{1},m_{2})$ is gauge-independent, it is easy to show that
Eq(\ref{S-matrix}) can be recast in the form
\begin{equation}
T(s,t,m_{1},m_{2})=
{\hat{T}}_{0}(t) + {\hat{T}}_{1}(t,m_{1})+
{\hat{T}}_{2}(t,m_{2})+
{\hat{T}}_{3}(s,t,m_{1},m_{2})~,
\label{S2-matrix}
\end{equation}
where the ${\hat{T}}_{i}$ ($i=0,1,2,3$) are {\sl individually}
$\xi$-independent.
  The propagator-like parts of vertex and box graphs
which enforce the gauge independence of $T_{0}(t)$,
 are called pinch parts.
They emerge every time a gluon propagator or an elementary
three-gluon vertex contributes a longitudinal $k_{\mu}$ to the original
graph's numerator. The action of such a term is
to trigger an elementary
Ward identity of the form
\be
\not\hspace*{-1.0mm}{k} =
(\not\hspace*{-1.0mm}{p}+\not\hspace*{-1.0mm}{k}-m)-
(\not\hspace*{-1.0mm}{p}-m)
\label{BasicPinch}
\ee
when it gets contracted with a $\gamma$ matrix.
The first term removes (pinches out) the
internal fermion propagator,
whereas the second vanishes on shell.
{}From the gauge-independent functions ${\hat{T}}_{i}$ ($i=0,1,2,3$)
one may now extract a gauge-independent
 effective gluon ($G$) self-energy
${\hat{\Pi}}_{\mu\nu}(q)$, gauge-independent
 $Gq_{i}{\bar{q}}_{i}$ vertices
${\hat{\Gamma}}_{\mu}^{(i)}$, and a gauge-independent box
$\hat{B}$, in the following way:
\begin{eqnarray}
&{\hat{T}}_{0}=g^{2}{\bar{u}}_{1}\gamma^{\mu}u_{1}
[(\frac{1}{q^{2}}){\hat{\Pi}}_{\mu\nu}(q)(\frac{1}{q^{2}})]
{\bar{u}}_{2}\gamma^{\nu}u_{2}~,\nonumber \\
&{\hat{T}}_{1}=g^{2}
{\bar{u}}_{1}{\hat{\Gamma}}_{\nu}^{(1)}u_{1}
(\frac{1}{q^{2}}){\bar{u}}_{2}\gamma^{\nu}u_{2}~,~~~~~~~~~~~~~~\\
&{\hat{T}}_{2}=g^{2}
{\bar{u}}_{1}\gamma^{\mu}u_{1}
(\frac{1}{q^{2}}){\bar{u}}_{2}
{\hat{\Gamma}}_{\nu}^{(2)}u_{2}~,~~~~~~~~~~~~~~\nonumber\\
&{\hat{T}}_{3}=\hat{B}~,~~~~~~~~~~~~~~~~~~~~~~~~~~~~~~
{}~~~~~~~~~~~~\nonumber
\label{DefOne}
\end{eqnarray}
where $u_{i}$ are the external spinors, and $g$ is the gauge coupling.
Since all hatted quantities in the above formula are
gauge-independent, their explicit
form may be calculated using any value of
the gauge-fixing parameter $\xi$, as long as one
properly
identifies and allots all relevant pinch contributions. The
choice $\xi=1$ simplifies the calculations significantly, since
it eliminates the longitudinal part of the gluon propagator.
Therefore, for $\xi=1$ the pinch contributions originate only
from momenta carried by
the elementary three-gluon vertex
The one-loop expression for ${\hat{\Pi}}_{\mu\nu}(q)$
is given by \cite{Bib} :
\begin{equation}
{\hat{\Pi}}_{\mu\nu}(q)= {\Pi}_{\mu\nu}^{(\xi=1)}(q)+
t_{\mu\nu}{{\Pi}^{P}}(q)~,
\label{Prop}
\end{equation}
where
\be
t_{\mu\nu}= (g_{\mu\nu}q^{2}-q_{\mu}q_{\nu})
\ee
and
\begin{equation}
{{\Pi}^{P}}(q)=-2ic_{a}g^2
\int_{n}\frac{1}{k^2(k+q)^2}~,
\label{som}
\end{equation}
where $\int_{n}\equiv\int \frac{d^{n}k}{{(2\pi)}^{n}}$ is the
dimensionally regularized loop integral, and $c_{a}$ is the
quadratic Casimir operator for the adjoint representation
[for $SU(N)$, $c_{a}=N$].
After integration and renormalization we find
\begin{equation}
{{\Pi}^{P}}(q) = -2c_{a}(\frac{g^2}{16\pi^{2}})
\ln(\frac{-q^{2}}{\mu^{2}})]~.
\end{equation}
Adding this to the Feynman-gauge proper self-energy
\begin{equation}
{\Pi}_{\mu\nu}^{(\xi=1)}(q)=-[\frac{5}{3}
c_{a}(\frac{g^2}{16\pi^{2}})\ln(\frac{-q^{2}}{\mu^{2}})]
t_{\mu\nu}~,
\end{equation}
we find for ${\hat{\Pi}}_{\mu\nu}(q)$
\begin{equation}
{\hat{\Pi}}_{\mu\nu}(q)=-bg^2\ln(\frac{-q^{2}}{\mu^{2}})t_{\mu\nu}~,
\label{RunnCoupl}
\end{equation}
where $b=\frac{11c_{a}}{48\pi^{2}}$ is the coefficient of
$-g^{3}$ in the usual $\beta$ function.

This procedure can be extended to an arbitrary $n$-point function;
of particular physical interest are the gauge-independent three and
four point functions
${\hat{\Gamma}}_{\mu\nu\alpha}$~\cite{C&P} and
$\hat{\Gamma}_{\mu\nu\alpha\beta}$~\cite{4g}, which at one-loop
satisfy the following {\it tree-level} Ward identities:
\bea
q_{1}^{\mu}\hat{\Gamma}_{\mu\nu\alpha}(q_{1},q_{2},q_{3})=&
  t_{\nu\alpha}(q_{2})\hat{d}^{-1}(q_{2}) -
 t_{\nu\alpha}(q_{3})\hat{d}^{-1}(q_{3}) \nonumber\\
q_{1}^{\mu}\hat{\Gamma}_{\mu\nu\alpha\beta}^{abcd} =&
f_{abp}\hat{\Gamma}_{\nu\alpha\beta}^{cdp}(q_1+q_2,q_3,q_4) + c.p.~,
\eea
where $\hat{d}={[q^{2}-\hat{\Pi}(q)]}^{-1}$ and
$f^{abc}$ the structure constants of the gauge group.

Finally, the generalization of the PT to the case of
non-conserved external currents is technically more involved
\cite{Bernd}.
The main reasons are the following:

{}~~(a)~~ The charged $W$
couples to fermions with different, non-vanishing masses $m_i,m_j \neq 0$,
 and
consequently the
elementary Ward identity of Eq.(\ref{BasicPinch}) gets modified to :
\be
k_{\mu}\gamma^{\mu} P_{L}\equiv \ksl P_{L} =
S_{i}^{-1}(p+k) P_{L} - P_{R}S_{j}^{-1}(p) + m_{i}P_{L} - m_{j}P_{R}
\label{BrokenPinch}
\ee
where
\be
P_{R,L} = \frac {1 \pm \gamma _{5}}{2}
\ee
are the chirality projection operators.
The first two terms of Eq(\ref{BrokenPinch}) will pinch and vanish on shell,
respectively, as they did before. But in addition, a term proportional to
$ m_{i}P_{L} - m_{j}P_{R}$
is left over.
In a general $R_{\xi}$ gauge such terms give rise to extra propagator
and vertex-like contributions, not present in the massless case.

{}~~(b)~~Additional graphs involving the ``unphysical" would-be Goldstone
bosons
$\chi$ and $\phi$, and physical Higgs $H$,
which do not couple to massless fermions,
must now be included.
Such graphs give rise to new pinch contributions,
even in the Feynman gauge, due to the momenta carried by
interaction vertices such as $\gamma\phi^{+}\phi^{-}$,
$Z\phi^{+}\phi^{-}$, $W^{+}\phi^{-}\chi$,  $HW^{+}\phi^{-}$,
 e.g. vertices with one vector gauge boson and two scalar bosons.

\section{The current algebra formulation of the pinch technique}

We now present an alternative
formulation of the PT introduced
in the context of the standard model~\cite{D&S}. In this
approach the interaction of gauge bosons with external fermions
is expressed in terms of
current correlation functions~\cite{Another Sirlin},
i.e. matrix elements of Fourier transforms
of time-ordered products of current operators.
This is particularly economical because these amplitudes automatically
include several closely related Feynman diagrams. When one of the current
operators is contracted with the appropriate four-momentum, a Ward identity
is triggered. The pinch part is then identified with the contributions
involving the equal-time commutators in the Ward identities, and therefore
involve amplitudes in which the number of current operators has been
decreased by one or more. A basic ingredient in this formulation are the
following equal-time commutators;
\begin{eqnarray}
&\delta(x_0-y_0)[J^{0}_{W}(x),J^{\mu}_{Z}(y)]=
 c^{2}J^{\mu}_{W}(x)\delta^{4}(x-y)~,\nonumber\\
&\delta(x_0-y_0)[J^{0}_{W}(x),J^{\mu\dagger}_{W}(y)]=
 - J^{\mu}_{3}(x)\delta^{4}(x-y)~,\\
&\delta(x_0-y_0)[J^{0}_{W}(x),J^{\mu}_{\gamma}(y)]=
 J^{\mu}_{W}(x)\delta^{4}(x-y)~,~~\nonumber\\
&\delta(x_0-y_0)[J^{0}_{V}(x),J^{\mu}_{V^{'}}(y)]= 0~,~~~~~~~
{}~~~~~~~~~~~~~\nonumber
\label{Commut2}
\end{eqnarray}
where $J_{3}^{\mu}\equiv 2(J_{Z}^{\mu}+s^{2}J_{\gamma}^{\mu})$
and $V,V^{'} \in \{ \gamma,Z \}$.
To demonstrate the method with an example, consider
the vertex $\Gamma_{\mu}$, where now the gauge
particles in the loop are $W$ instead of
gluons and the incoming and outgoing fermions are massless.
It can be written as follows (with $\xi=1$):
\begin{equation}
\Gamma_{\mu}=\int \frac{d^{4}k}{{2\pi}^4}
\Gamma_{\mu\alpha\beta}(q,k,-k-q)\int d^{4}x e^{ikx}
<f|T^{*}[J^{\alpha\dagger}_{W}(x)J^{\beta}_{W}(0)]|i>~.
\label{Papous}
\end{equation}
When an appropriate momentum, say $k_{\alpha}$,
 from the vertex is pushed into the integral over
$dx$, it gets transformed into a covariant derivative
 $\frac{d}{dx_{\alpha}}$ acting on the time ordered product
$<f|T^{*}[J^{\alpha\dagger}_{W}(x)J^{\beta}_{W}(0)]|i>$.
After using current
conservation and differentiating the
$\theta$-function terms, implicit in the definition of
the $T^{*}$ product, we end up with the left-hand side
 of the second of Eq(22).
So, the contribution of each such term is proportional to the
matrix
element of a single current operator,
 namely $<f|J_{3}^{\mu}|i>$; this
is precisely the pinch part. Calling $\Gamma_{\mu}^{P}$
 the total pinch contribution from the
$\Gamma_{\mu}$ of Eq(\ref{Papous}), we find that
\begin{equation}
\Gamma_{\mu}^{P}= -g^{3}cI_{WW}(Q^2)<f|J_{3}^{\mu}|i>~,
\label{PinchPapou}
\end{equation}
where
\begin{equation}
I_{ij}(q)= i\int_{n}\frac{1}
{(k^{2}-M_{i}^{2})[{(k+q)}^{2}-M_{j}^{2}]}~.
\label{IntegralIWW}
\end{equation}
Obviously, the integral in Eq(\ref{IntegralIWW}) is the generalization
of the QCD expression Eq(\ref{som})
to the case of massive gauge bosons.

\section{Gauge--invariant gauge boson vertices
 and their Ward identities}
We consider
the S-matrix element for the process
\begin{equation}
e^-  + \nu  +  e^- \to
e^- + e^-  +  {\overline \nu}     ~~.
\label{process}
\end{equation}
and isolate the part $T(q,p_1,p_2)$ of the S--matrix which depends
only on the momentum transfers $q$, $p_1$, and $p_2$.
The tree-level vector-boson propagator $\Delta_{\mu\nu}^{i}(q)$
 in the $R_{\xi}$ gauges is given by
\be
\Delta^{\mu\nu}_{i}(q,\xi_i)= \frac{1}{q^{2}-M^{2}_{i}}
[g^{\mu\nu} - (1-\xi_{i})\frac{q^{\mu}q^{\nu}}
{q^{2}-\xi_{i}M^{2}_{i}}]~~,
\label{Prop1}
\ee
with $i=W,Z,\gamma$, and $M_{\gamma}=0$.
Its inverse
 $\Delta_{i}^{-1}(q,\xi_i)^{\mu\nu}$
is given by
\be
\Delta_{i}^{-1}(q,\xi)^{\mu\nu}= (q^{2}-M^{2}_{i})g^{\mu\nu}
- q^{\mu}q^{\nu}+ \frac{1}{\xi_{i}}q^{\mu}q^{\nu}~~.
\label{InvProp}
\ee
The propagators $\Delta _{s}(q,\xi_i)$ of the
unphysical (would--be) Goldstone bosons are
given by
\be
\Delta_{s}(q,\xi_{i})=\frac{-1}{q^{2}-\xi_{i} M^{2}_{i}} ~~,
\label{Gold}
\ee
with $(s,i) = (\phi,W)$ or $(\chi,Z)$
and explicitly depend on $\xi_{i}$. On the other hand,
the propagators of the fermions (quarks and leptons), as well as the
propagator
of the physical Higgs particle
are $\xi_{i}$-independent at tree-level.

Since the final result (with pinch contributions included)
is gauge-independent, we choose to work in the Feynman gauge
($\xi_{i}=1$); this particular gauge simplifies the calculations
because it removes all longitudinal parts from the
tree-level gauge boson propagators. So, pinch contributions can
only originate from appropriate momenta furnished by the
tree--level gauge boson vertices.
Applying the pinch technique algorithm we isolate all
vertex--like parts contained
in the box diagrams and allot them to the usual vertex
graphs.
The final expressions for
one loop gauge-independent trilinear gauge boson vertices are :
\bea
\frac{1}{g^3s}
\hGa^{\gamma W^{-}W^{+}} _{\mu \alpha \beta} & =&
\Gamma^{\gamma W^-W^+}_{\mu \alpha \beta} |_{\xi _i =1}
 + q^{2}~B_{\mu\alpha \beta}
 +U_W^{-1}(p_1)^{\rho}_{\alpha} B^+_{\mu \rho \beta}
+U_W^{-1}(p_2)^{\rho}_{\beta} B^-_{\mu \alpha \rho}
\nonumber \\
&& - 2 \Omega\Gamma_{\mu\alpha\beta}
+ p_{2 \beta} g_{\mu \alpha} ~ {\cal M}^-
  +p_{1 \alpha} g_{\mu \beta}~  {\cal M}^+ ~~,
\label{gigWW}
\eea
\bea
\frac{1}{g^3c}
\hGa^{ZW^{-}W^{+}} _{\mu \alpha \beta} & =&
\Gamma^{Z W^-W^+}_{\mu \alpha \beta} |_{\xi _i =1}
+ U_Z^{-1}(q)^{\rho}_{\mu} B_{\rho \alpha \beta}
 +U_W^{-1}(p_1)^{\rho}_{\alpha} B^+_{\mu \rho \beta}\nonumber \\
  && +U_W^{-1}(p_2)^{\rho}_{\beta} B^-_{\mu \alpha \rho}
-  2 \Omega\Gamma_{\mu\alpha\beta}
+ q_{\mu} g_{\alpha \beta}~M_Z^2~ {\cal M}\nonumber \\
&&+ p_{2 \beta} g_{\mu \alpha} ~M_W^2~ {\cal M}^-
  + p_{1 \alpha} g_{\mu \beta}~M_W^2~  {\cal M}^+
\label{giZWW} ~~,
\eea
where
\be
\Omega=
I_{WW}(q)+s^2I_{W\gamma}(p_1)+c^2I_{WZ}(p_1)+s^2I_{W\gamma}(p_2)
+c^2I_{WZ}(p_2)~,
\ee
and
\be
{\cal M}^-(q,p_1,p_2) = ~\frac {s^2}{c^2}
 J_{WW\gamma} +
\frac{1-2s^2}{2c^2} ~J_{WWZ}  + \frac{1}{2} ~J_{WWH}
+\frac{1}{2c^2} ~J_{ZHW}~
\label{M-} ~,
\ee
with
\be
J_{ABC}
=\int_{n}
\frac{1}
{\left[(k+p_1)^2-M_A^2\right]\left[(k-p_2)^2-M_B^2\right]
\left[k^2-M_C^2\right]}~,
\ee
and the property
\be
{\cal M}^+(q,p_1,p_2) = -{\cal M}^-(q,p_2,p_1) ~.
\ee

The gauge-independent vertices satisfy the following simple Ward
identities (WI), relating them to the
$W$ self energy and $\chi WW$ vertex constructed also via the PT :
\be
q^\mu \hGa ^{ZW^-W^+} _{\mu \alpha \beta}
+ iM_Z \hGa ^{\chi W^-W^+} _{\alpha \beta}
 =  gc ~\left[ \hPi ^W _{\alpha \beta} (1) - \hPi ^W _{\alpha \beta} (2)
\right]~~,
\label{qZWW}
\ee
\be
q^\mu \hGa ^{\gamma W^-W^+} _{\mu \alpha \beta}
 =  gs ~\left[ \hPi ^W _{\alpha \beta} (1) - \hPi ^W _{\alpha \beta} (2)
\right] ~~.
\label{qGWW}
\ee
These WI
 are the one--loop generalizations of the respective {\it tree level} WI;
their validity is
crucial for the gauge independence of the S--matrix.
It is important to emphasize that they make no
reference to ghost terms, unlike the corresponding Slavnov-Taylor
identities satisfied by the conventional, gauge--dependent vertices.

For the case of {\it on--shell}
{}~$W$s one sets $p_1^2=p_2^2=M^2_W$ and neglects all
terms proportional to $p_{1\alpha}$ and $p_{2 \beta}$,
as well as the left over pinch terms of the $W$ legs. Then the
$\gamma WW$ vertex reduces to the form
\be
\frac{1}{g^3s}
\hGa^{\gamma W^{-}W^{+}} _{\mu \alpha \beta}  =
\Gamma^{\gamma W^-W^+}_{\mu \alpha \beta} |_{\xi _i =1}
 + q^{2}~B_{\mu\alpha \beta}(q,p_1,p_2)
 - 2 \Gamma_{\mu\alpha\beta}
I_{WW}(q)~~.
\label{gigWW2}
\ee
This is of course the same answer one obtains by applying the PT
{\it directly} to the S--matrix of $e^{+}e^{-} \rightarrow W^{+}W^{-}$.
Thus for the form factors ${\Delta\kappa}_{\gamma},~{\Delta Q}_{\gamma}$
the only function we need is $B_{\mu\alpha \beta}$, given below
\be
g^2 B_{\mu\alpha \beta} = \sum_{V=\gamma Z} g_{V}^2
\int_{n}
\frac{iR_{\alpha\beta\mu}}
{\left[(k+p_1)^2-M_W^2\right]\left[(k-p_2)^2-M_W^2\right]
\left[k^2-M_V^2\right]}
{}~,
\ee
with
\be
R_{\alpha\beta\mu}=
g_{\alpha \beta} ~(k-\frac{3}{2}(p_1-p_2))_{\mu}
-g_{\alpha \mu}~(3k+2q)_{\beta}
-g_{\beta \mu}~(3k-2q)_{\alpha}~.
\ee

\section{Magnetic dipole and electric quadrupole form factors for the $W$}

Having constructed the
gauge-independent $\gamma WW$ vertex we proceed to extract
its contributions to the magnetic dipole and electric quadrupole form
factors of the $W$.
We use carets to denote the gauge independent one--loop contributions.
Clearly,
\be
\hat{\Delta}\kappa_{\gamma} = \Delta\kappa_{\gamma}^{(\xi=1)}
+ \Delta\kappa_{\gamma}^{P}~,
\label{fullkappa}
\ee
and
\be
\hat{\Delta}Q_{\gamma} = \Delta Q_{\gamma}^{(\xi=1)}+  \Delta Q_{\gamma}^{P}
{}~,
\label{fullQ}
\ee
where $\Delta Q_{\gamma}^{(\xi=1)}$ and $\Delta Q_{\gamma}^{(\xi=1)}$
are the contributions of the usual vertex diagrams in
the Feynman gauge \cite{Lahanas}, whereas
$\Delta Q_{\gamma}^P$ and $\Delta Q_{\gamma}^P$ the analogous
contributions from the pinch parts.
The task of actually calculating
${\hat{\Delta}\kappa}_{\gamma}$ and ${\hat{\Delta} Q}_{\gamma}$ is greatly
facilitated by the fact that the quantities
${\Delta\kappa}_{\gamma}^{(\xi=1)}$ and ${\Delta Q}_{\gamma}^{(\xi=1)}$
have already been calculated in \cite{Lahanas}. It must be emphasized
however that the expression for ${\Delta\kappa}_{\gamma}^{(\xi=1)}$
(but not  ${\Delta Q}_{\gamma}^{(\xi=1)}$)
is infrared {\sl divergent} for $Q^2 \not = 0$ due to the presence of the
following double integral over the Feynman parameters (t,a), given in
Eq.(26) of \cite{Lahanas}:
\bea
R &= -(\frac{\alpha_{\gamma}}{\pi})\frac{Q^{2}}{M^{2}_{W}}
\int^{1}_{0} da \int^{1}_{0}
\frac{dtt}{t^{2}-t^{2}(1-a)a(\frac{4Q^2}{M^2_{W}})}\nonumber\\
&= -(\frac{\alpha}{2\pi})\frac{Q^{2}}{M^{2}_{W}}
\int^{1}_{0}
 \frac{da}{1-(1-a)a(\frac{4Q^2}{M^2_{W}}}) \int^{1}_{0}
\frac{dt}{t}
\label{BadGuy}
\eea
By performing the momentum integration in $B_{\mu\alpha\beta}$, we find for
$p_1^2=p_2^2=M_W^2$
\be
B_{\mu\alpha\beta}= -\frac{Q^2}{8\pi^2 M^2_W}\sum_{V=\gamma, Z} g^2_{V}
\int_{0}^{1}da \int_{0}^{1}(2tdt) \frac{F_{\mu\alpha\beta}}{L_{V}^{2}}~,
\ee
where
\be
F_{\mu\alpha\beta}
= 2(\frac{3}{2} + at)g_{\alpha\beta}{\Delta}_{\mu}+
2(3at+2)[g_{\alpha\mu}Q_{\beta}-g_{\beta\mu}Q_{\alpha}]~,
\label{F}
\ee
and
\be
L^{2}_{V} = t^{2}-t^{2}a(1-a)(\frac{4Q^{2}}{M^{2}_{W}}) +
 (1-t)\frac{M^{2}_{V}}{M^{2}_{W}}~,
\ee
from which immediately follows that
\be
a_{1}^{P}(Q^{2})= -\frac{1}{2}\frac{Q^{2}}{M^{2}_{W}}
\sum_{V}\frac{\alpha_{V}}{\pi} \int_{0}^{1}da \int_{0}^{1}(2tdt)
\frac{2(\frac{3}{2} + at)}{L^{2}_{V}}
\ee
and
\be
a_{2}^{P}(Q^{2})= -\frac{1}{2} \frac{Q^{2}}{M^{2}_{W}}
\sum_{V} \frac{\alpha_{V}}{\pi}
 \int_{0}^{1}da \int_{0}^{1}(2tdt)
\frac{2(2+3at)}{L^{2}_{V}}~,
\ee
and since there is no term proportional to $\Delta_{\mu}Q_{\alpha}Q_{\beta}$,
\be
a_{3}^{P}(Q^{2})=0~.
\ee
Therefore,
\be
{\Delta\kappa}_{\gamma}^{P}= -\frac{1}{2} \frac{Q^{2}}{M^{2}_{W}}
\sum_{V} \frac{\alpha_{V}}{\pi} \int_{0}^{1}da \int_{0}^{1}(2tdt)
\frac{(at-1)}{L^{2}_{V}}~,
\label{ligoulaki}
\ee
and
\be
{\Delta Q}_{\gamma}^{P} = 0 ~~~.
\ee
It is important to notice that even though ${\Delta Q}_{\gamma}^{P} = 0$
{\sl both} $\mu_{W}$ and $Q_{W}$ will assume values different
than those predicted in the
$\xi=1$ gauge.
Indeed,
even though the value of $\lambda_{\gamma}$ does not change, the value of
$\kappa_{\gamma}$ changes, and this change affects both
$\mu_{W}$ and $Q_{W}$ through Eq(\ref{dipole}) and Eq(\ref{quad}).
In the expression given in Eq(\ref{ligoulaki})
the first term (for V=Z) is infrared finite (since $M_{Z} \not = 0$),
whereas the second term (for $V=\gamma$) is infrared divergent, since
$M_{\gamma}=0$. Calling this second term $\Theta$ we have
\be
\Theta= -\frac{1}{2}
(\frac{\alpha_{\gamma}}{\pi})\frac{Q^{2}}{M^{2}_{W}}
 \int_{0}^{1}da \int_{0}^{1}dt
\frac{2t(at-1)}{t^{2}[1-a(1-a)\frac{4Q^{2}}{M^{2}_{W}}]}~,
\label{Theta}
\ee
which can be rewritten as
\be
\Theta=  -R - (\frac{\alpha_{\gamma}}{\pi})\frac{Q^{2}}{M^{2}_{W}}
\int_{0}^{1}da\frac{a}{1-a(1-a)\frac{4Q^{2}}{M^{2}_{W}}}~,
\label{AlmostThere}
\ee
where $R$ is the infrared divergent integral
defined in Eq(\ref{BadGuy}). On the other hand,
the second term in Eq(\ref{AlmostThere})
is infrared finite.
Clearly, including the first term of Eq(\ref{AlmostThere}) in the value of
${\hat{\Delta}\kappa}_{\gamma}$ {\sl exactly} cancels the infrared divergent
contribution of Eq(\ref{BadGuy}),
thus giving rise to an infrared finite expression
for ${\hat{\Delta}\kappa}_{\gamma}$.
So, after the infrared divergent part of Eq(\ref{Theta}) is cancelled,
${\Delta\kappa}_{\gamma}^{P}$ is given by the following expression:
\be
{\Delta\kappa}_{\gamma}^{P}= \Theta_{\gamma} + \Theta_{Z}~,
\label{AntePali}
\ee
with $\Theta_{\gamma}$ the second term in Eq(\ref{AlmostThere}),
and $\Theta_{Z}$
the second term in Eq(\ref{ligoulaki}), namely
\be
\Theta_{\gamma}=
- (\frac{\alpha_{\gamma}}{\pi})\frac{Q^{2}}{M^{2}_{W}}
\int_{0}^{1}da\frac{a}{1-a(1-a)\frac{4Q^{2}}{M^{2}_{W}}}~,
\label{ThetaGamma}
\ee
and
\be
\Theta_{Z}=
 -\frac{Q^{2}}{M^{2}_{W}}
 (\frac{\alpha_{Z}}{\pi}) \int_{0}^{1}da \int_{0}^{1}dt
\frac{t(at-1)}{L^{2}_{Z}}~,
\label{ThetaZeta}
\ee
and from Eq(\ref{fullkappa})
\be
{\hat{\Delta}\kappa}_{\gamma} = {[{\Delta\kappa}_{\gamma}^{(\xi=1)}]}_{if} +
\Theta_{\gamma} + \Theta_{Z}~,
\label{End}
\ee
where the subscript $(if)$ in the first term of the R.H.S. indicates
that the contribution from the $\xi=1$ gauge is now genuinely
infrared finite.
Finally, the magnetic dipole moment $\mu_{W}$ and electric quadrupole
moment $Q_{W}$ are given by
\be
\mu_{W}= \frac{e}{2M_{W}}(2+{\hat{\Delta}\kappa}_{\gamma})
\label{Finaldipole}
\ee
and
\be
Q_{W}= -\frac{e}{M^{2}_{W}}(1+{\hat{\Delta}\kappa}_{\gamma}
+ 2{\hat{\Delta} Q}_{\gamma})~.
\label{Finalquad}
\ee
Both ${\Delta Q}_{\gamma}^{(\xi=1)}$
 and ${\Delta\kappa}_{\gamma}^{(\xi=1)}$
have been computed numerically in \cite{Lahanas}.
We now proceed to compute the integrals in Eq(\ref{ThetaGamma})
 and Eq(\ref{ThetaZeta}), which
determine ${\Delta\kappa}_{\gamma}^{P}$.
It is elementary to evaluate $\Theta_{\gamma}$.
 Setting
$\Theta_{\gamma}=-(\frac{\alpha_{\gamma}}{\pi}){\hat{\Theta}}_{\gamma}$
we have:
\bea
{\hat{\Theta}}_{\gamma}&= \frac{2}{\Delta}
[arctg(\frac{1}{\Delta})- arctg(\frac{-1}{\Delta})],
{}~~~~ Q^{2}<M_{W}^{2}\nonumber\\
&= -4 , ~~~~ Q^{2}=M_{W}^{2}~~~~~~~~~~~~~~~~~~~~~~~~~~~~~\\
&= \frac{2}{\Delta}\ln[\frac{|\Delta - 1|}{\Delta + 1}],
{}~~~~Q^{2}>M_{W}^{2}~~~~~~~~~~~~~~~~~~\nonumber
\eea
for space-like $Q^{2}$,
where $\Delta= \sqrt{|\frac{M^{2}_{W}}{Q^{2}}-1|}$,
and
\be
\hat{\Theta}_{\gamma} =
\frac{2}{\Delta}\ln[\frac{|\Delta - 1|}{\Delta + 1}] ~~
\ee
for time-like $Q^{2}$, where
$\Delta= \sqrt{\frac{M^{2}_{W}}{|Q^{2}|}+1}$.

The double integral
$\Theta_{Z}$ can in principle be expressed in a closed form in terms
of Spence functions [see for example
\cite{Veltman}],
but this is of limited usefulness for our present calculation.
Instead, we evaluated this integral numerically.
We used the same values for the constants
appearing in our calculations as in \cite{Lahanas},
namely $\alpha_{\gamma}=\frac{1}{128}$, $M_{W}=80.6 GeV$,
$M_{Z}=91.1 GeV$ and $s=0.23$ .

The result of the computation is very interesting.
$\Delta\kappa_{\gamma}^{P}$, which originates from pinching box
diagrams, furnishes exactly the contributions needed to
restore the unitarity of the final answer.
Indeed, as the authors of \cite{Lahanas} emphasized,
$\Delta\kappa_{\gamma}^{(\xi = 1)}$ is by itself not a gauge
invariant object in the limit $Q^2 \rightarrow\infty$,
where the local $SU(2)\times U(1)$ symmetry is restored.
For large values of $Q^2$, $\Delta\kappa_{\gamma}^{P}$
is nearly equal in magnitude and opposite in sign to
$\Delta\kappa_{\gamma}^{P}$. Therefore, when according to
Eq(\ref{fullkappa}) and Eq(\ref{fullQ}) both contributions are added,
$\hat{\Delta}\kappa_{\gamma} \rightarrow 0$
as $Q^2 \rightarrow \pm\infty$.
Clearly, the inclusion of the pinch parts from the
box graphs is {\it crucial} for restoring the
good asymptotic behavior of the $W$ form factors.

\eject
\section{Conclusions}
We presented a study of the structure of
trilinear gauge boson vertices
in the context of the standard model.
Using the S-matrix pinch technique
gauge-independent $\gamma WW$ and $ZWW$ vertices were
constructed to one-loop order, with all three incoming momenta
off-shell.
These vertices
satisfy naive QED--like Ward identities,
which relate them to the gauge independent $W$ self-energy,
which were also obtained via the pinch technique.
The tree-level Ward identities are to be contrasted with the
complicated Ward identities satisfied by the conventionally
defined gauge-dependent vertices; in particular, no
ghost terms need be included.
Finally, when the appropriate Lorentz structures are
extracted, these vertices
give rise to
gauge-independent, infrared finite, and asymptotically
well-behaved
magnetic dipole and electric quadrupole form factors for the $W$,
which
can, at least in principle, be promoted to physical
observables.
It would be interesting to determine how these quantities could be
directly extracted from future $e^{+}e^{-}$ experiments.

\section {Acknowledgment}
I thank
J.~M.~Cornwall, K.~Hagiwara, A.~Lahanas, K.~Philippides,
R.~Peccei,
and D.~Zeppenfeld for useful discussions, and E.~Karagiannis
for his warm hospitality during my visit in Los Angeles.
This work was supported by the National Science Foundation under
Grant No.PHY-9017585.

\end{document}